\documentclass{article}

\def\mytitle#1{\setcounter{equation}{0}
\setcounter{footnote}{0}
\begin{flushleft}\Large\textbf{#1}\end{flushleft}
\vspace{0.25cm}}
\def\myname#1{\leftline{{\large #1}}\vspace{-0.13cm}}
\def\myplace#1#2{\small\begin{flushleft}\textit{#1}\\
\texttt{#2}\end{flushleft}}
\newenvironment{contribution}{\normalsize\noindent}{}
\def\myclassification#1{\small\noindent
Pacs no : 97.60.Lf ; 04.70.Dy
       #1\vspace{0.5cm}}
\usepackage{graphicx}% Include figure files
\begin{document}

\mytitle{Hawking radiation as tunnelling from the trapping horizon of a general non-static spherically symmetric space-time}

\vskip0.2cm \myname{Subenoy
Chakraborty\footnote{schakraborty@math.jdvu.ac.in}} \vskip0.2cm
\myname{Nairwita Mazumder\footnote{nairwita15@gmail.com}}
 \vskip0.2cm
\myname{Ritabrata Biswas\footnote{biswas.ritabrata@gmail.com}}
\vskip0.2cm

\myplace{Department of Mathematics, Jadavpur University,
Kolkata-32, India.}{} \vskip0.2cm

\myclassification{}

\begin{abstract}
Using tunnelling approach, Hawking radiation is derived for a
general class of non-static spherically symmetric space time. The
standard tunnelling rate formula is obtained using the unified
first law of thermodynamics on the trapping horizon and Hawking
temperature is measured by an Kodama observer inside the horizon.
\end{abstract}

\begin{contribution}
\end{contribution}
\section{Introduction}
One of the greatest challenges in theoretical physics still today is to develop a theory of quantum gravity i.e. to formulate a theory of gravity within the frame work of quantum mechanics. A major achievement towards this goal is the discovery of Hawking radiation \cite{Hawking1} as it is generally speculated that black hole (BH) thermodynamics has a key role in this unification process. Hawking showed that a BH radiates and hence should behave like a perfect thermodynamical system. Hawking \cite{Hawking1} was able to show the radiation from the BH considering Bogolyubov coefficients between asymptotic in and out states in a  collapse geometry and subsequently, in collaboration with Gibbons \cite{Gibbons1, Birrell1} he showed the thermal radiation in Euclidean quantum gravity considering the periodicity in Euclidean time to resolve the conical singularity. Also there are other attempts to derive Hawking radiation \cite{Hartle1, Christensen1} . However, in recent past, a semi classical tunneling method \cite{Kraus1, Kraus2, Parikh1, Parikh2, Angheben1, Srinivasan1, Shankaranarayanan1, Shankaranarayanan2}  attracts many people due to simplicity of calculations and for a new physical inside. In this semi classical approach, the key points are the consideration of energy conservation in tunneling of a thin shell from the hole and the background metric is allowed to fluctuate in the tunneling process. Further, a general analysis \cite{Hu1, Sarkar1, Pilling1, Zhang1} of this tunneling approach with the first law of thermodynamics for horizons ($dE=Tds-pdV$) shows that the tunneling rate $\Gamma\sim\ exp\left\{\Delta S\right\}$, ($\Delta S$ is the variation of entropy).\\

For static BH, there exist an event horizon and Hawking temperature is well defined for this global concept. But difficulty for dynamical BH where event horizon may not exist. Recently, using tunneling method Hayward et al \cite{Hayward1} were able to define Hawking temperature locally on the trapping horizon of dynamical BH.\\

On the other hand, this tunneling approach was also applied to the universe as a thermodynamical system. Sarkar et al \cite{Sarkar1} derived hawking radiation through tunneling mechanism for general asymptotically flat spherically symmetric space times and demonstrated the intimate relation between the tunneling approach and the first law of thermodynamics. They have concluded that their results are valid for any gravity theory where first law of thermodynamics holds on the horizon. Very recently, Cai et al \cite{Cai1} has showed that the existence of Hawking radiation associated with the locally defined apparent horizon of the FRW universe using the tunneling approach and an observer with the Kodama vector \cite{Kodama1, Minamitsuji1,Racz1} inside the horizon measures the Hawking temperature. Subsequently, Jiang et al \cite{Jiang1} extend the work of Sarkar et al \cite{Sarkar1} to FRW universe and are able to derive the tunneling formula $\Gamma\sim exp\left\{\Delta S\right\}$, from the inner relationship of tunnelling method with the unified first law of thermodynamics ($dE_{n}=TdS+WdV$)though the principle of conservation of energy. in the present paper, we go further one step forward to generalize the work of Sarkar et. al. \cite{Sarkar1} for a general non-static spherically symmetric space-time. Using the unified first law, the standard tunnelling rate formula is derived at the trapping horizon, for a Kodama observer inside the horizon.\\\\

\section{General non static spherically symmetric space time and tunnelling formation :}
In general, a spherically symmetric metric can be written in local form as
\begin{equation}\label{1}
ds^{2}=q_{ij}\left(x^{i}\right)dx^{i}dx^{j}+R^{2}(x^{i})d\Omega^{2}
\end{equation}
where $i,~j$ can take values $0$ and $1$, the two dimensional
metric
\begin{contribution}
\begin{equation}\label{2}
d\gamma^{2}=q_{ij}\left(x^{i}\right)dx^{i}dx^{j}
\end{equation}
is called the normal metric having coordinates $x^{i}$and the area radius $R\left(x^{i}\right)$ may be considered as a scalar field in the normal 2-D space. Further, vanishing of the scalar
\begin{equation}\label{3}
h(x)=q^{ij}\left(x\right)\partial_{i}R\partial_{j}R
\end{equation}
identify the dynamical trapping horizon i.e., at the trapping horizon ($H_{T}$) we have
\begin{equation}\label{4}
h(x)|_{H_{T}}=0~~~~~,~~~~\partial_{i}h|_{H_{T}}\neq 0
\end{equation}
According to Hayward \cite{Hayward2} the expansions ($\theta_{+}$) of the null future directed (outgoing) congruence normal to a section of the horizon vanishes. On the other hand, corresponding to "incoming rays" thereis another null congruence with expansion scalar $\theta_{-}$. Further, if $\theta_{-}<0$ and ${\cal L}_{n}\theta_{+}<0$ along the horizon then it is classified as future, outer type. In the present work we shall consider such a trapping horizon.

Now, the dynamical surface gravity \cite{Hayward3} associated with the dynamical horizon is given by
\begin{equation}\label{5}
\kappa_{T}=\frac{1}{2}Box_{q}R_{|_{H_{T}}}=\frac{1}{2\sqrt{-q}}\partial_{i}\left(\sqrt{-q}q^{ij}\partial_{j}R\right)|_{H_{T}}
\end{equation}\\
For dynamical spherically symmetric space time, the Kodama vector field $\kappa^{\mu}$ is characterised by $\left(\kappa^{\mu}G_{\mu\nu}\right)^{;\nu}=0$.
For the present choice of the metric (given by equation (\ref{1}))
\begin{equation}\label{6}
k^{i}=\frac{1}{\sqrt{-q}}\epsilon^{ij}\partial_{j}R~~~~,~~~~k^{\theta}=0=k^{\phi}
\end{equation}
where $\epsilon^{ij}$ is the volume from associated with the 2D
space $\left(x^{0},~x^{1}\right)$. the Kodama vector is a
generalisation of the time-like Killing vector for static
space-time and it gives a preferred flow of time. Also similar to
the usual Killing identity the Kodama vector satisfies on the
horizon
\begin{equation}\label{7}
k^{\mu}\nabla_{[\nu, K^{\mu}]}=\pm\kappa_{T}\kappa_{\nu}
\end{equation}
The trapping horizon is further classified as outer , degenerate
or inner if $\kappa_{T}>0,~~\kappa_{T}=0$ or $\kappa_{T}<0$
respectively.

To generalize the tunnelling approach for this dynamical
spherically symmetric space-time we must take into account of (i)
application energy conservation in such a general set up and (ii)
unified first law of thermodynamics. In general  relativity, it is
difficult to define the concept of energy, so for energy
conservation one has to depend on some heuristic arguments. The
unified first law of thermodynamics has the form

\begin{equation}\label{8}
dE_{h}=TdS+WdV
\end{equation}

where $\rho$, $p$ are the usual energy density and thermodynamic
pressure of the perfect fluid and the work density $W$ has the
expression $W=\frac{1}{2}\left(\rho-p\right)$. It is to be noted
that this identity can be interpreted as the energy conservation
relation when horizon is virtually displaced normal to itself. On
the other hand, from the point of view of standard thermodynamics
\cite{Jiang1}, the unified first law is a connecting relation
between two quasi-static equilibrium states of a system having
infinitesimal differences $dV$, $dS$ and $dE_{h}$ of the extensive
variables volume, entropy and energy respectively but the
intensive variables namely temperature $T$, pressure $p$ and
density $\rho$ remain same,i.e., the quasi static states may be
considered as the spherically symmetric solutions of the Einstein
equations having same matter source only the horizon radius
differs infinitesimally. However, from tunnelling view point when
some matters tunnels out or in across the horizon, there is a
change of the energy of the whole space time and as a result the
energy of the shell can be determined. Also the Hamiltonian of the
tunnelling particles is related to the imaginary part of the
action in the s-wave WKB approximation \cite{Jiang1}.

Introducing Painleve type coordinates the metric (\ref{1}) for general spherically symmetric space time can be written as

\begin{equation}\label{9}
ds^{2}=d\gamma^{2}+r^{2}d\Omega^{2}
\end{equation}

where

\begin{equation}\label{10}
d\gamma^{2}=-E(r,~t)dt^{2}+2F(r,~t)dtdr+G(r,~t)dr^{2}
\end{equation}
with $F\neq0$.

Then
\begin{equation}\label{11}
h=\gamma^{ij}\partial_{i}r\partial_{j}r=\gamma^{rr}\left(t,~r\right)=\frac{E}{EG+F^{2}}
\end{equation}
So trapping horizon is located at $E_{H}=0~~(F_{H}\neq0)$
The Kodama vector and the dynamical surface gravity at the horizon are given by
\begin{equation}\label{12}
K=\left(\frac{1}{\sqrt{F^{2}+EG}}, \hat{0}\right)=\left(\frac{1}{F}, \hat{0}\right)
\end{equation}
and
\begin{equation}\label{13}
\kappa_{H}=\left[\frac{1}{2F^{3}}\left(E'F-\frac{1}{2}\dot{E}G\right)\right]_{H}
\end{equation}
The radial null geodesic can be expressed as
\begin{equation}\label{14}
\frac{dr}{dt}=\frac{1}{G}\left[-F\pm \sqrt{F^{2}+EG}\right]
\end{equation}
where outgoing (oringoing) null geodesics correspond to $+$ve (or
$-$ve) sign. In the present context as the observer is inside the
trapping horizon so we consider only incoming mode.

So near the trapping horizon we have
\begin{equation}\label{15}
\frac{dr}{dt}=-\kappa_{H}\left(r-r_{H}\right)F\left(1-\frac{1}{2}\frac{\dot{E}G}{E'F}\right)^{-1}
\end{equation}
Let us now consider an ingoing positive energy particle which
crosses the horizon inwards from $r_{i}$ to $r_{f}$. Then the
imaginary part of the action for the corresponding s-wave can be
written as
\begin{equation}\label{16}
Im S=Im \int_{r_{i}}^{r_{f}}p_{r}dr=Im \int_{r_{i}}^{r_{f}}\int_{0}^{p_{r}}dp_{r}'dr
\end{equation}
To change the integration variable from momentum to energy $y$ we
shall use the Hamilton's equation of motion  $\dot{r}=\frac{d{\cal
H}}{dp_{r}}$. Thus changing the order of integration we have
\begin{equation}\label{17}
Im S=Im\int \int_{r_{i}}^{r_{f}}\frac{dr}{\dot{r}}d{\cal H}
\end{equation}
Using $\dot{r}$ from the radial null geodesic near the horizon the
integration over the radial coordinate can be evaluated by
approapriate choice of the contour surrounding the pole at
$r=r_{{\cal H}}$ and we obtain

\begin{equation}\label{18}
Im S=\int_{{\cal H}_{i}}^{{\cal H}_{f}}\frac{d{\cal H}}{2T}\frac{1}{F}\left(1-\frac{1}{2}\frac{\dot{E}G}{E'F}\right)
\end{equation}

where Hawking temperature $T$ appears due to elimination of surface gravity using the Bekenstein relation $\frac{|\kappa|}{2\pi}=T$
before evaluating the aboveintegral over the Hamiltonian we note that due to explicit time-dependence of the present dynamical system, the hamiltonian is no longer equivalent to the total energy of the system. However, to obtain the relation between the Hamiltonian and the total energy of the system we follow the Hayward's work \cite{Hayward1, Cai1, Akbar1} as follows :
There are two conserved currents associated with general spherical dynamical system - one is the usual Kodama vector $K^{\mu}$ and the other is the energy momentum density $j^{a}=T_{b}^{a}K^{b}$, along the Kodama vector. The associate conserved charge with $j^{a}$ is nothing but thr Misner-sharp energy, i.e., $E=-\int_{\sigma} j^{a}d\sigma_{a}$, where $d\sigma_{a}$ is the oriented volume form of the space like hypersurface $\sigma_{a}$. It is easy to see that the total energy bounded by the trapping horizon is $E_{H}=\frac{r_{H}}{2}$. Thus the total energy of the space time can be writen as

\begin{equation}\label{19}
E_{T}=\frac{r_{H}}{2}-\int_{\sigma}T_{b}^{a}K^{b}d\sigma_{a}
\end{equation}
where integral ranges from the horizon to infinity.

Now during the tunnelling process, physical parameters like mass,
charge etc change and as a result the radius of the trapping
horizon changes. Suppose in the initial and final states of the
tunnelling process the trspping horizon has radius $r_{H}$ and
$r_{H}+dr_{H}$, while matter source remains same . Thus according
to a Kodama observer the energy change due to the tunnelling
process contributes to the shell and hencefrom energy conservation
\begin{eqnarray}\label{20}
\nonumber d{\cal H}=E^{f}_{T}\left(r_{H}+dr_{H}\right)-E_{T}^{i}(r_{H})~~~~~~~~~~~~~~~~~~~~~~~~~~~~\\
\nonumber ~~~~~~~~~~~~~~~~~~~~~~~~ =\frac{\delta  r_{H}}{2}-\left[\int_{r_{H}+dr_{H}}^{\infty}T_{b}^{a}K^{b}d\sigma_{a}-
\int_{r_{H}}^{\infty}T_{b}^{a}K^{b}d\sigma_{a}\right]~~\\
~=dE_{H}-\rho dV~~~~~~~~~~~~~~~~~~~~~~~~~~~~~~~~~~~~~~~~~~~~~
\end{eqnarray}
This energy difference is measured by a kodama observer inside the trapping horizon so near the trapping horizon
\begin{equation}\label{21}
d{\cal H}=\frac{dH}{\frac{1}{F}}=FdH
\end{equation}
Using (\ref{20}) and (\ref{21}) in (\ref{18}) we obtain
\begin{eqnarray}\label{22}
\nonumber ImS =\int_{H_{i}}^{H_{f}}\frac{dH}{2T}\left(1-\frac{1}{2}\frac{\dot{E}G}{E'F}\right)\\
=\int\frac{\left(dE_{H}-\rho dV\right)}{2T}\left(1-\frac{1}{2}\frac{\dot{E}G}{E'F}\right)
\end{eqnarray}
Using the conservation equation and the total energy $E_{H}$ inside the trapping horizon as
$$E_{H}=\rho V~~~~with~~~~V=\frac{4}{3}\pi r_{H}^{3}$$
the above equation (\ref{22}) can be simplified as
\begin{equation}\label{23}
Im S=\int\frac{\left(dE_{H}-WdV\right)}{2T}
\end{equation}
using the unified first law (\ref{8}), we have the desired result
$$Im S=\int \frac{dS}{2}$$
Hence the semiclassical tunnelling rate for the present general spherically symmetric space-time model of the universe is given by
$$\Gamma\sim exp\left\{-2ImS\right\}=exp\left\{\int_{S_{i}}^{S_{f}}dS\right\}=exp\left\{-\Delta S\right\}~~~~~,~~~where~~\Delta S=S_{f}-S_{i}>0$$
Which is welknown result for tunnelling probability. Thus the tunnelling interpretation of Hawking radiation is intimately depend on unified first law of thermodynamics. this close association is due to the fact that both depends on the energy conservation.\\\\

\section{Conclusion}
The present work is an extension of the work in ref
\cite{Sarkar1}. As we have considered general spherically
symmetric non-static space-time, so we have trapping horizon and
time like Killing vector is replaced by the Kodama vector.
Consequently, the first law of thermodynamics is generalised by
the unified first law of thermodynamics. However, we have obtained
identical tunnelling rate, as both tunnelling process and unified
first law of thermodynamics are closely associated with the energy
conservation. The Hawking temperature for the general dynamical
system is measured by an kodama observer inside the trapping
horizon.\\\\
 {\bf Acknowledgement :}

NM wants to thank CSIR, India for awarding JRF. RB is thankful to West Bengal State Govt for awarding JRF.

\end{contribution}

\end{document}